# FPGA Digital Dice using Pseudo Random Number Generator

A PREPRINT

Michael Lim Kee Hian, Ten Wei Lin, Zachary Wu Xuan, Stephanie-Ann Loy, Maoyang Xiang, T. Hui Teo*
Engineering Product Development
Singapore University of Technology and Design
*Corresponding Author: tthui@sutd.edu.sg
These authors contributed equally to this work.

**ABSTRACT**

The goal of this project is to design a digital dice that displays dice numbers in real-time. The number is generated by a pseudo-random number generator (PRNG) using XORshift algorithm that is implemented in Verilog HDL on an FPGA. The digital dice is equipped with tilt sensor, display, power management circuit, and rechargeable battery hosted in a 3D printed dice casing. By shaking the digital dice, the tilt sensor signal produces a seed for the PRNG. This digital dice demonstrates a set of possible random numbers of 2, 4, 6, 8, 10, 12, 20, 100 that simulate the number of dice sides. The kit is named SUTDicey.

# I. Introduction and Implementation

SUTDicey is a hardware module (Figure 1) that provides an interactive solution for simulating dice rolls, offering users the ability to select and roll various types of dice, from 2-sided to 100-sided. Integrated with a tilt sensor and pseudorandom number generator (PRNG), it enables the real-time display of dice roll results on a 7-segment display, delivering an engaging experience for gaming and other applications.

The BOM is summarized in Table 1 for reference. Each Verilog HDL module's behaviour is also given for those who are interested in making similar digital dice.

## RNG Computation

The XORshift algorithm used was adapted in this design, as this algorithm has the significant advantage of speed for real-time application, [1,2]. An XORshift algorithm works by taking the exclusive or (XOR) of a number with a shifted version of itself multiple times, before feeding back into the input again.

The use of a 32-bit seed in the XORshift method makes the randomness seemingly close to a true random number generator (TRPG). A UART connection was used to plot a statistical distribution and evaluate the randomness of the generated XORshift outputs.

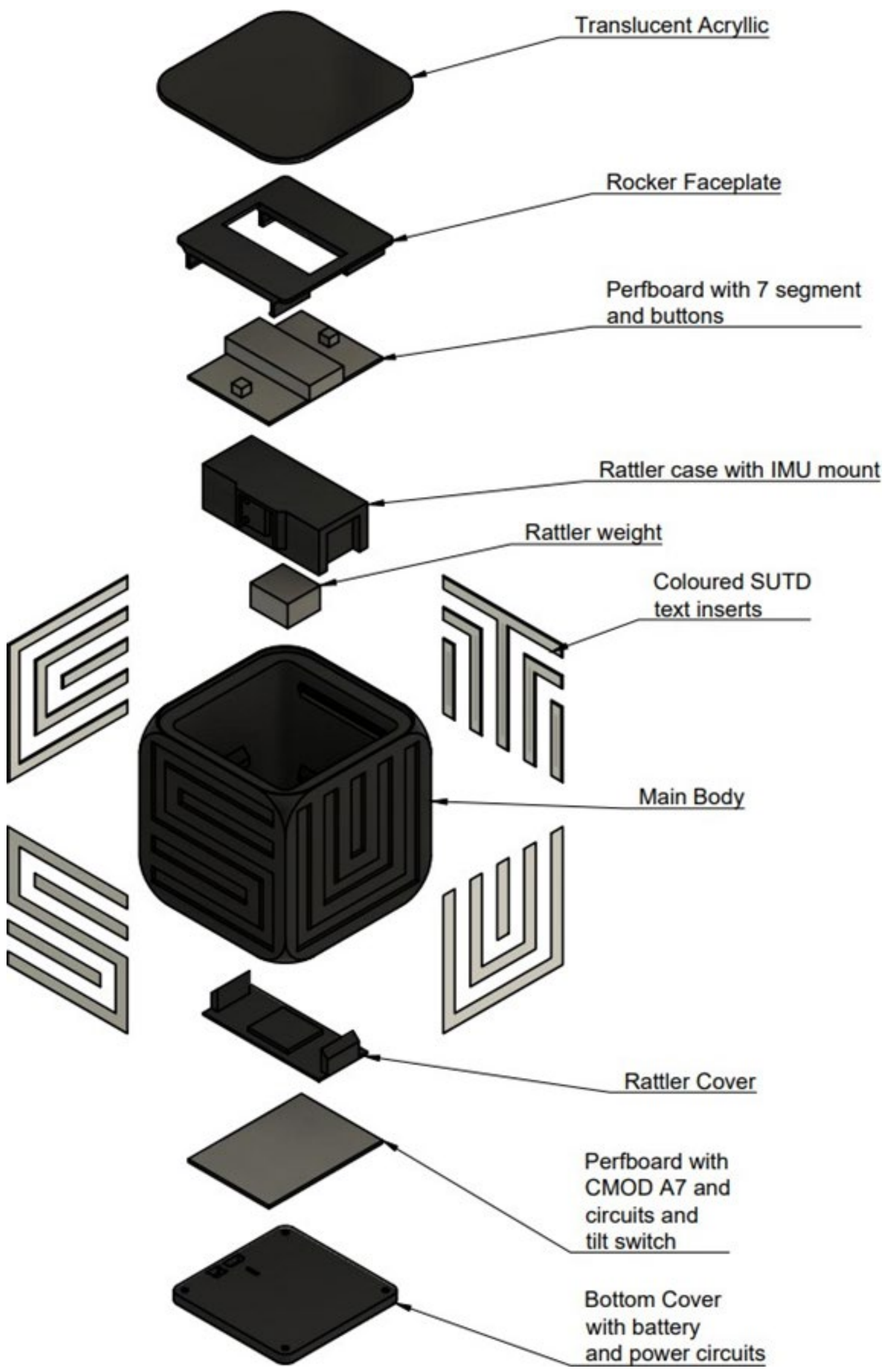

*Figure 1: Exploded view of SUTDicey.*

*Table 1:  Bill Of Materials*

| | |
|---|---|
| CMOD A7 35T | Accelerometer adxl335 |
| Tilt Switch sw-520d | Steel Weight for rattler |
| 3D printed parts (see Figure 1) | 3x 9.1k ohm resistors |
| USB-C PD trigger (5v) | 3x 10k ohm resistors |
| Dc-Dc battery charge/discharge module (5v) | 3x 6mm tactile pushbutton |
| 3.3v dc-dc step down synchronous buck regulator D24V10F3 | 3.7v LiPo battery |
| 7 segment display 5643BH | Translucent Acrylic Faceplate |
| 4x 3Mx4 heat-set threaded inserts | 4x 3Mx8 hex head screws |
| 2x perforated board 70mmx50mm | |

Various electronics components such as CMOD-A7, battery, DC-DC Charge Discharge Integrated Module, Step-down voltage regulators, power push button, up/down push button, tilt switch and resistors (9.1k & 10k) were used and interconnected based on Figure 2.

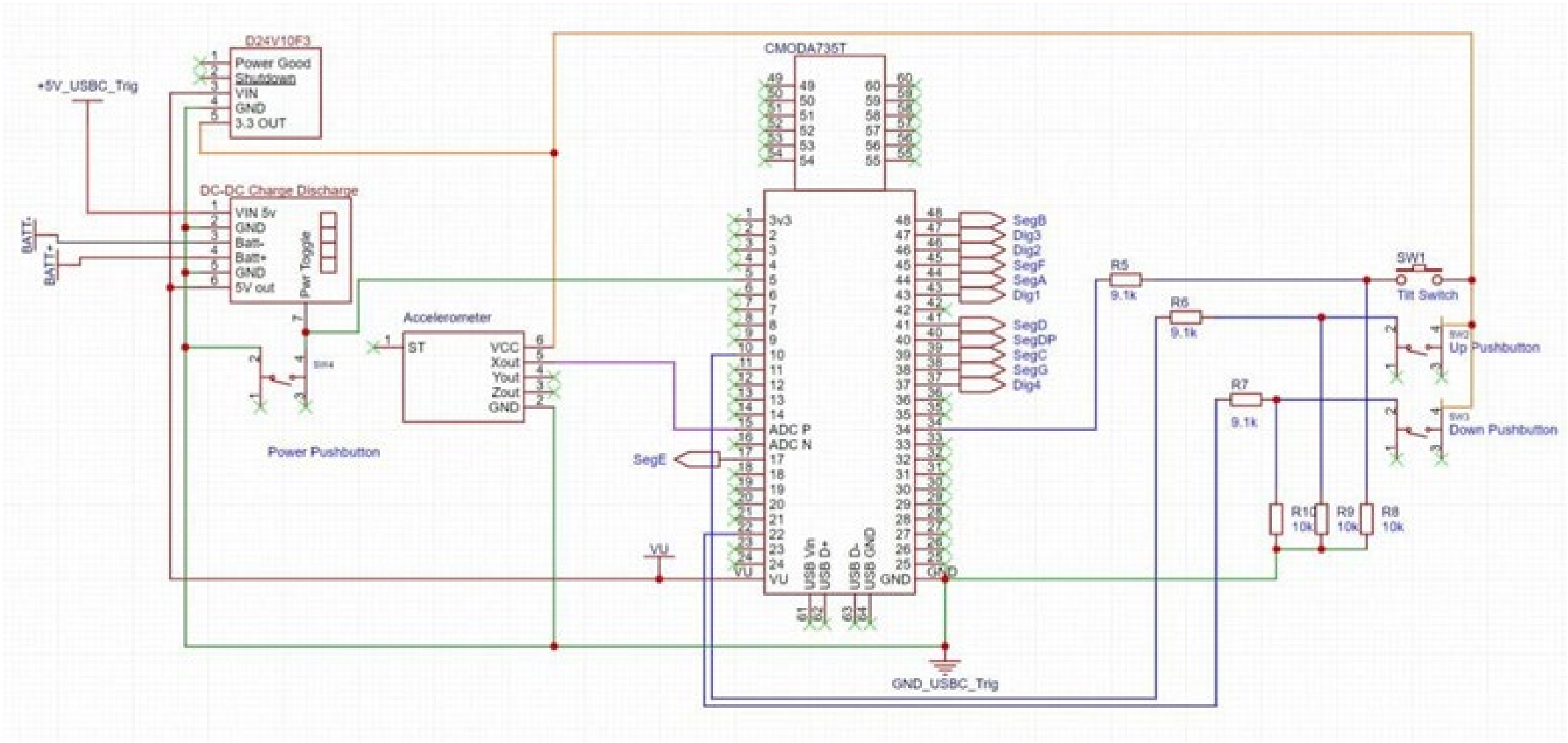

*Figure 2: Wiring Circuit for SUTDicey.*

## Constraint File

The constraint file is intended for the CmodA7 rev. B FPGA board. It defines pin assignments and I/O standards for various peripherals and interfaces.

To implement the file in the project simply:

1. Uncomment lines corresponding to used pins
2. Rename the ports (in each line, after get_ports) according to the top-level signal names in the project.

Furthermore, it's worth noting that the file also contains a commented-out line. If uncommented, it permits combinational logic loops with a warning severity level.

## Top Module

This section summarises the functionality and behavioural description of each block in the top module Verilog file.

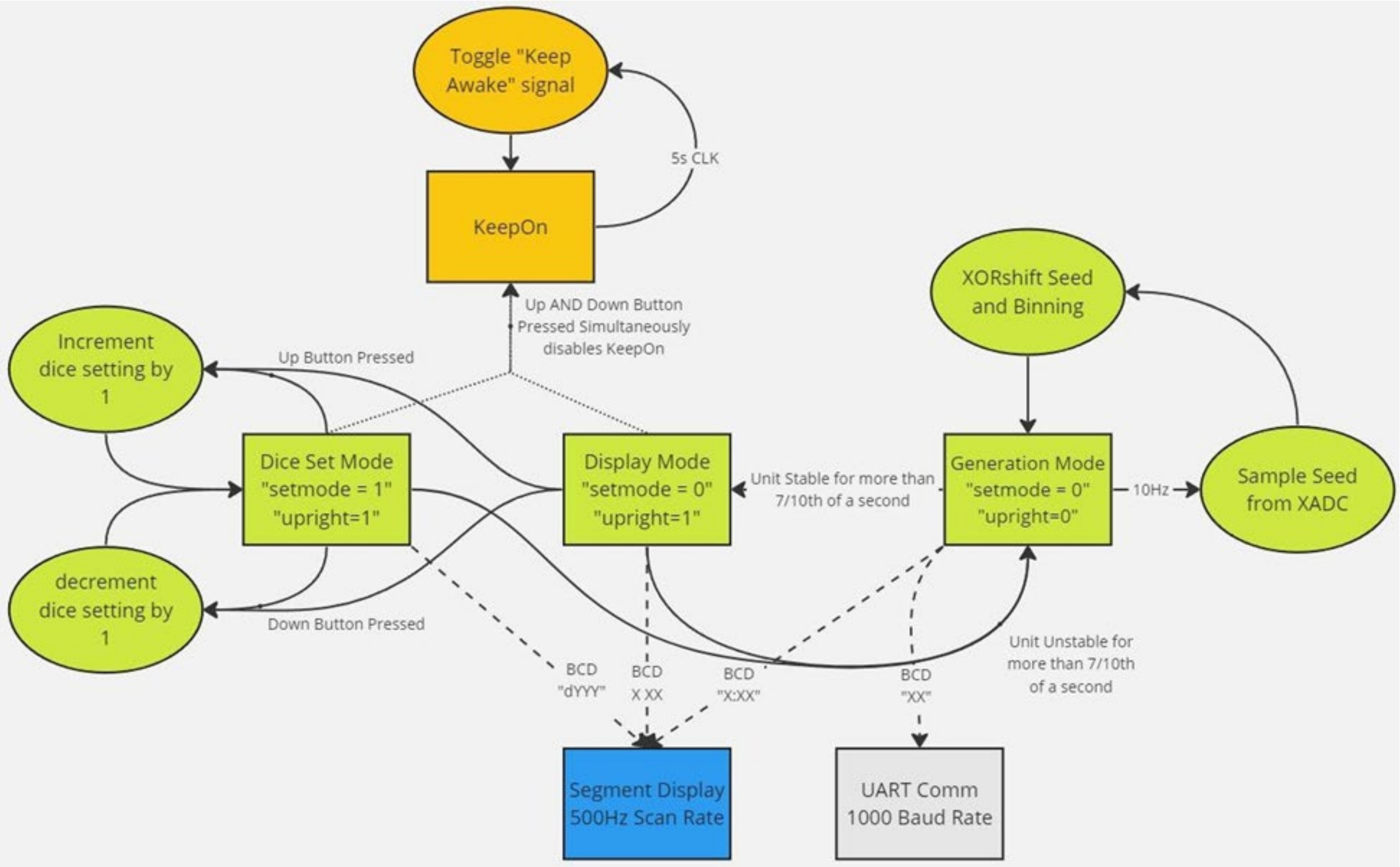

*Figure 3: Simplified State Flow Chart (Solid lines represent state changes, broken/dotted lines represent information flow Colours represent separate loops).*

# Tilt Detection

Continuously monitors the state of a tilt sensor to determine whether a dice unit is upright and stable.

Behavioural Description:

1. Initialization and Reset Handling:
   - Upon system initialization or a reset signal (rstn), the system resets its internal state.
   - sumtilt is reset to 0, indicating no prior detection of the tilt sensor being in an "on" state.
   - The tiltlog shift register is cleared, ensuring a fresh start for storing tilt sensor readings.
   - The upright flag is set to 0, indicating that the dice unit is not currently detected as upright.
2. Continuous Monitoring and Upright Determination:
   - The system continuously monitors the state of the tilt sensor by sampling it at each positive edge of the CLK10Hz signal.
   - The current state of the tilt sensor is stored in the tiltlog shift register, replacing the oldest recorded state.
   - The sum of the bits in the tiltlog register is calculated, representing the duration the tilt sensor has been in the "on" state over a specific period (approximately 1 second).
   - If the sum of the tiltlog register exceeds a predetermined threshold (7), indicating that the tilt sensor has been consistently in the "on" state for a certain duration (about 7/10th of a second), the system sets the upright flag to 1.
   - If the sum of the tiltlog register is below the threshold, the upright flag remains at 0, indicating that the dice unit is not considered upright and stable.

# Dice Selection and Display

This block manages the control flow of the dice controller, allowing users to interact with it to set the dice value (2,4,6,8,10,12,20,100-sided dice) and mode based on button presses.

Behavioural Description:

- On a clock edge or reset, it initializes/reset various signals and variables.
- It updates the states of the "up" and "down" buttons (btnUr and btnDr).
- It checks if the dice is upright (upright) and processes button presses accordingly:
  - If both the "up" and "down" buttons are pressed simultaneously, it disables the keep awake feature (keepon).
  - If only the "up" button is pressed, it enters the set mode and cycles through dice values (dselect).
  - If only the "down" button is pressed, it enters the set mode and cycles through dice values in reverse.
- It updates the displayed value on the segment display (thou_set, huns_set, tens_set, ones_set) based on the selected mode (dselect).
- If the dice is not upright, it exits the set mode.

## Dice Selection and Display

This block manages the control flow of the dice controller, allowing users to interact with it to set the dice value (2,4,6,8,10,12,20,100 sided dice) and mode based on button presses.

Behavioural Description:

- On a clock edge or reset, it initializes/reset various signals and variables.
- It updates the states of the "up" and "down" buttons (btnUr and btnDr).
- It checks if the dice is upright (upright) and processes button presses accordingly:
  - If both the "up" and "down" buttons are pressed simultaneously, it disables the keep awake feature (keepon).
  - If only the "up" button is pressed, it enters the set mode and cycles through dice values (dselect).
  - If only the "down" button is pressed, it enters the set mode and cycles through dice values in reverse.
- It updates the displayed value on the segment display (thou_set, huns_set, tens_set, ones_set) based on the selected mode (dselect).
- If the dice is not upright, it exits the set mode.

## XADC Analog Input to 32-bit Seed Generator

This module converts XADC analogue input to a 32-bit seed using a shift register. It accumulates ADC data on each positive clock edge, generating (Segment_Data) as the seed PRNG.

Behavioural Description:

- On the positive edge of the clock signal (CLK10Hz), or upon a negative reset (rstn), the module either resets the seed value to zero or shifts in new ADC data.
- If a reset occurs, the seed value is reset to zero.
- Otherwise, the module shifts the current seed value left by 16 bits and assigns the lower 16 bits to the newly acquired ADC data.
- This results in the generation of a 32-bit value in (Segment_Data) as a seed for PRNG.

## Segmented Display Processing

Processes the raw RNG values generated by the PRNG algorithm and prepares them for display on a segmented display.

Behavioural Description:

- The raw RNG values (rand) are generated by the pseudo-random number generator (PRNG) algorithm discussed earlier.
- These raw RNG values are fed into the display processing logic to convert them into BCD (Binary Coded Decimal) format for display.
- The RNG values are modulo-ed by a dice value to normalize them into the desired range for the dice game. Essentially involves converting a wide range of RNG values into a smaller range suitable for a dice roll (e.g., converting from a 32-bit range to a 1-6 range for a six-sided dice).
- The processed RNG values are then converted to BCD format for display on the segmented display.

## Segment Display Control

The 7-segment display can be set manually or generated randomly.

Behavioural Description:

- Four variables ones_bcd, tens_bcd, huns_bcd, and thou_bcd, each 4 bits wide, representing the BCD (Binary Coded Decimal) values for ones, tens, hundreds, and thousands places.
- If setmode is true, it sets the BCD values directly from thou_set, huns_set, tens_set, and ones_set. If setmode is false, it sets the BCD values based on some random values (thou_rand, huns_rand, tens_rand, ones_rand), with some conditions.
- BCD values are assigned to bcd_tim in the appropriate order for display.

## Battery Module

It toggles a pin every 5 seconds to prevent the module from shutting down due to low current draw from FPGA. The keep-awake function can be disabled by pressing both up and down buttons simultaneously.

## UART Interface (Communication)

Controls the rate of message transmission and outputs data representing the tens and ones value of a dice roll through UART

Behavioural Description:

- uart_ready toggles on each positive clock edge, controlling the message transmission rate.
- Data is sent over UART when uart_ready is high and uart_valid is asserted.
- The UART interface operates asynchronously, transmitting data via uart_rxd_out.
- The display cycles through the digits at a rate of 500 Hz (clk500hz).

## Segment Module

Controls a 4-digit 7-segment display, sequentially showing each digit of a BCD number.

Behavioural Description:

- The module sequentially displays each digit of the BCD number on the 7-segment display.
- An internal register (an_r) determines which digit is currently being displayed, cycling through each digit in sequence.
- The BCD number to be displayed is selected based on the value of an_r.
- The BCD number is converted into the corresponding 7-segment code and stored in an internal register (segment_r).
- On reset (rstn), all displays are turned off and overwritten with “dddd”.

## UART TX Module

The UART transmitter module is designed to transmit serial data asynchronously to a receiver. It operates based on a finite state machine (FSM) to control the transmission process. The module handles data transfer, start and stop bit insertion, and optionally, parity bit generation for error detection.

Behavioural Description:

- The module uses an FSM to manage the transmission process. The FSM transitions between states including IDLE, START, TRANSFER, and STOP based on various conditions such as readiness of the receiver (ap_ready).
- When data is ready (ap_ready), the module starts the transmission process by sending a start bit (FSM_STAR). It then proceeds to transmit each bit of the data (data) followed by an optional parity bit (parity) and a stop bit (FSM_STOP).
- The module generates output signals (ap_valid and tx) to indicate the validity of the transmitted data and the actual serial data stream, respectively.
- Synchronous reset (ap_rstn) is used to initialize the module to the IDLE state.
- Optionally, the module can perform parity checking (FSM_PARI) to ensure data integrity, although the implementation for this is currently commented out.

## Random Number Generation for Dice Roll

A pseudorandom number generator (PRNG) is used to generate random numbers for the dice rolls. The PRNG algorithm takes seed values obtained from a XADC and produces pseudorandom values, which are used for generating the dice roll results.

Behavioural Description:

- On each rising edge of the system clock (sysclk), the PRNG computes the next pseudo-random number.
- It uses the previous pseudo-random number (stored in rand_reg) to compute the next value.
- The computation involves bitwise XOR and shifting operations.
- The result is stored in a register (rand_reg) synchronously to avoid combinatorial loop issues.
- The pseudo-random number is outputted as rand for further use in the FPGA design.

```verilog
reg [31:0] temp;
reg [31:0] temp2;
reg [31:0] rand;

//Pseudo-Random XOR shift register algorithm with seed value supplied by XADC
always@(sysclk)begin
temp = Segment_data ^ Segment_data >>7;
temp2 = temp ^ temp << 9;
rand = temp2 ^ temp2 >> 13;
end
```

The generated PRNG numbers are plotted in Figure 4, and Figure 5, which show the randomness.

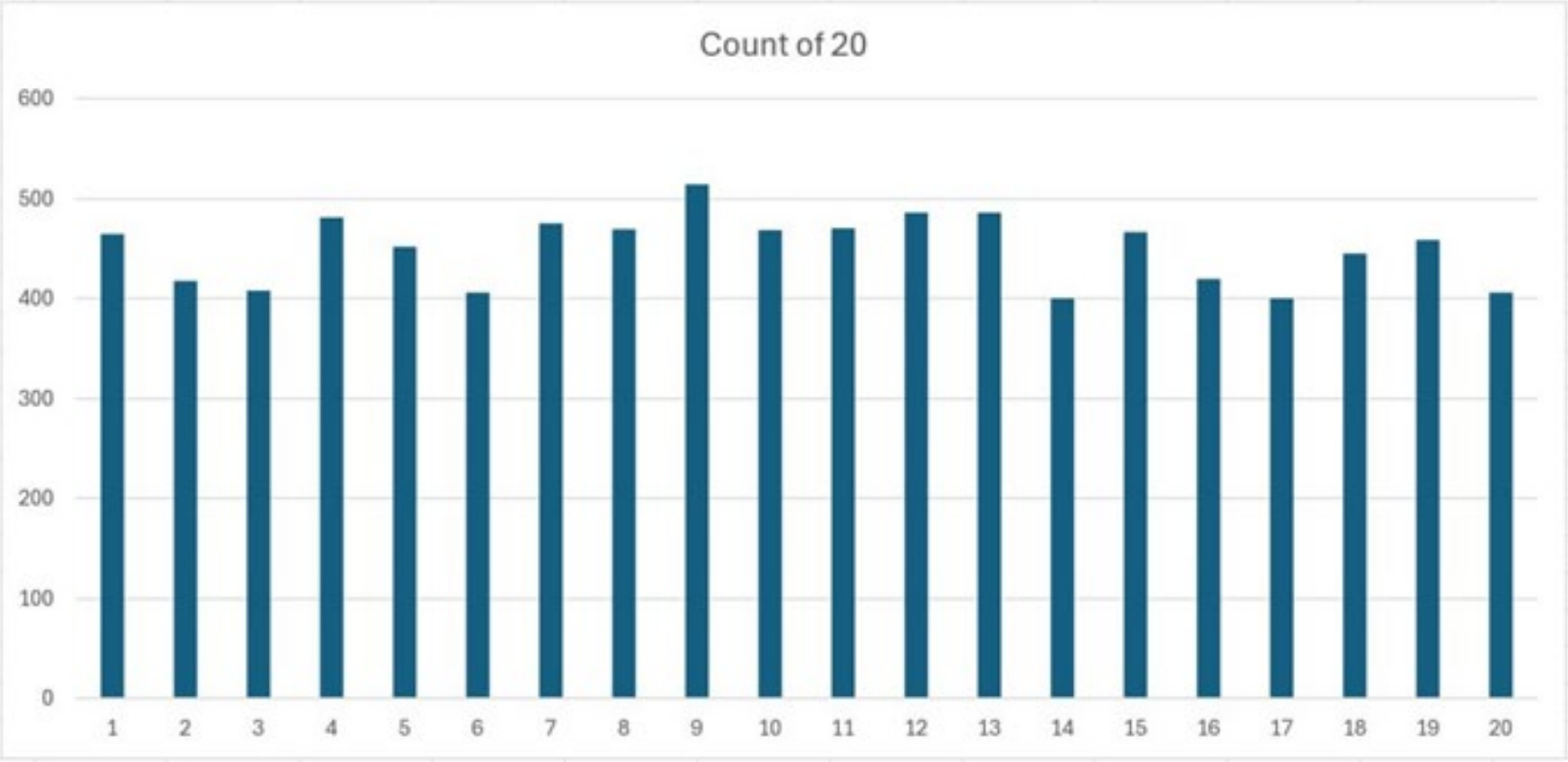

*Figure 4: Statistical distribution for 20-sided dice.*

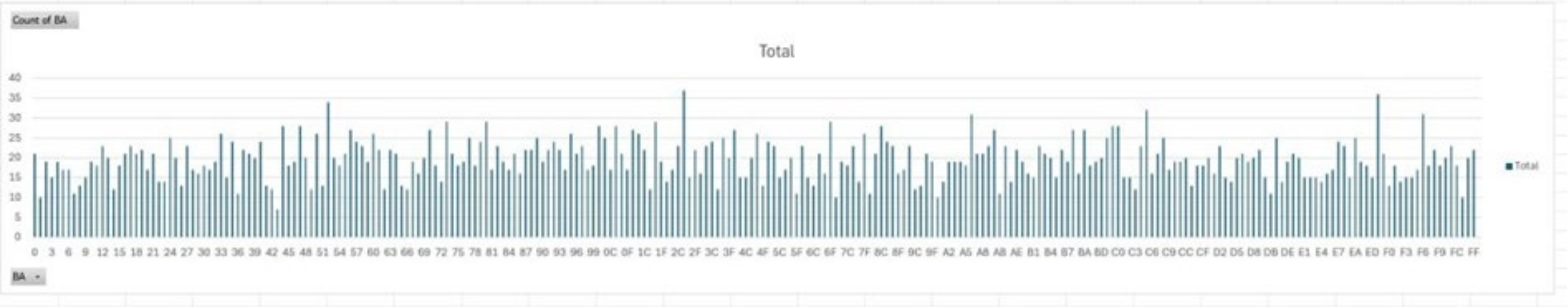

*Figure 5: Statistical distribution before RNG value is converted to the chosen dice range.*

# II. Operation of SUTDicey

This section is dedicated to user guides for using the dice. The key steps are given below.

1. Power on SUTDicey by pressing the recessed button on the bottom of the unit, a red light should show in the slot nearby.
    a. If a red light does not show, the unit is probably out of power, please charge it with a USB C cable (5 V mode)
2. Place the SUTDicey unit upright, with the smooth acrylic side facing up, it will take a few seconds to "boot" up. When numerals are displayed through the acrylic face, it has completed booting.
3. Press above or below on the acrylic face to change between the d2, d4, d6, d8, d10, d12, d20 and d100-sided dice.
    a. Pressing both up and down buttons at the same time while the dice is upright will initiate shutdown which will occur after around 15-30 seconds.
4. To roll the selected dice, tilt upside down or shake the SUTDicey unit for more than one second, a colon in the middle of the display will light to show that the unit is not stable and is "tossing/scrambling" the dice. The display numbers will also scramble and randomize.
5. Place or hold the unit upright for more than a second to "roll" the die. The colon in the middle of the display will disappear once the die is "stable" and has "settled" on a final value. Be careful not to shake or upset the SUTDicey unit as it will roll the dice again if it detects enough movement.
6. Shaking or tilting the SUTDicey unit will roll the last type of dice selected.

## Summary

A complete design of a digital dice using PRNG is demonstrated in the work. More work to be done in improving the implementation of the PRNG algorithm in FPGA is required for better randomness that meeting actual dice's behaviour.

# Acknowledgments

We would like to thank SUTD-ZJU IDEA Visiting Professor Grant (SUTD-ZJU (VP) 202103, and SUTD-ZJU Thematic Research Grant (SUTD-ZJU (TR) 202204), for supporting this work.

# Appendix

```
module top_module(
        input sysclk,
        input tilt,
        input btnU,
        input btnD,
        input [1:0] btn,
        output onpin,
        output [6:0] seg,// 7-Segment - Segment[6:0];
        output dp,      // 7-Segment - Segment-DP;
        output [3:0] an, // 7-Segment - Common Anode;
        output [1:0] led,
        input vp_in,
        input vn_in,
        input [1:0] xa_n,
        input [1:0] xa_p,
        output uart_rxd_out
);

reg CLK1000Hz,CLK1500Hz,CLK500Hz,CLK10Hz,CLK5s;
wire [15:0] bcd_tim;
wire rstn;
reg keepon;
reg onsig;
reg clk5;

assign rstn=~btn[0];
assign led[1]=tilt;
assign led[0]=clk5;

reg [12:0] CLK_CNTER_1000Hz;
reg [12:0] CLK_CNTER_1500Hz;
reg [13:0] CLK_CNTER_500Hz;
reg [19:0] CLK_CNTER_10Hz;
reg [24:0] CLK_CNTER_5s;

//Generate 1000Hz CLK;
```

```
always@(posedge sysclk, negedge rstn)begin
    if(!rstn) begin
        CLK_CNTER_1000Hz<=13'd0;
        CLK1000Hz <= 1'b0;
    end
    else begin
        if(CLK_CNTER_1000Hz == 13'd6000-1'b1) begin
            CLK1000Hz <= ~ CLK1000Hz;
            CLK_CNTER_1000Hz <= 13'd0;
        end
        else CLK_CNTER_1000Hz <= CLK_CNTER_1000Hz + 1'b1;
    end
end

//Generate 1500Hz CLK;
always@(posedge sysclk, negedge rstn)begin
    if(!rstn) begin
        CLK_CNTER_1500Hz<=13'd0;
        CLK1500Hz <= 1'b0;
    end
    else begin
        if(CLK_CNTER_1500Hz == 13'd4000-1'b1) begin
            CLK1500Hz <= ~ CLK1500Hz;
            CLK_CNTER_1500Hz <= 13'd0;
        end
        else CLK_CNTER_1500Hz <= CLK_CNTER_1500Hz + 1'b1;
    end
end

//Generate 5s CLK;
always@(posedge sysclk, negedge rstn)begin
    if(!rstn) begin
        CLK_CNTER_5s<=25'd0;
        CLK5s<= 1'b0;
    end
    else begin
        if(CLK_CNTER_5s == 25'd30001200-1'b1) begin
            CLK5s <= ~ CLK5s;
            CLK_CNTER_5s <= 25'd0;
        end
        else CLK_CNTER_5s <= CLK_CNTER_5s + 1'b1;
    end
end

//Generate 500Hz CLK;
always@(posedge sysclk, negedge rstn) begin
    if(!rstn) begin
        CLK_CNTER_500Hz<=14'h0000;
        CLK500Hz <= 1'b0;
    end
    else begin
        if(CLK_CNTER_500Hz == 14'd12_000-1'b1) begin
            CLK500Hz <= ~ CLK500Hz;
            CLK_CNTER_500Hz <= 14'h0000;
        end
        else CLK_CNTER_500Hz <= CLK_CNTER_500Hz + 1'b1;
    end
end

//Generate 10Hz CLK;
always@(posedge sysclk, negedge rstn)begin
    if(!rstn) begin
        CLK_CNTER_10Hz<=20'd0;
        CLK10Hz <= 1'b0;
    end
    else begin
        if(CLK_CNTER_10Hz == 20'd600024-1'b1) begin
            CLK10Hz <= ~ CLK10Hz;
            CLK_CNTER_10Hz <= 20'd0;
        end
        else CLK_CNTER_10Hz <= CLK_CNTER_10Hz + 1'b1;
    end
end

```

```
//XADC IP block setup
reg [31:0] Segment_data;
wire enable;                        //enable into the xadc to continuosly get data out
reg [6:0] Address_in = 7'h14;       //Adress of register in XADC drp corresponding to data
wire ready;                         //XADC port that declares when data is ready to be taken
wire [15:0] ADC_data;                   //XADC data

xadc_wiz_0 xadc_u0
(
    .daddr_in(Address_in),           // Address bus for the dynamic reconfiguration port
    .dclk_in(sysclk),                // Clock input for the dynamic reconfiguration port
    .den_in(enable),                 // Enable Signal for the dynamic reconfiguration port
    .di_in(0),                       // Input data bus for the dynamic reconfiguration port
    .dwe_in(0),                      // Write Enable for the dynamic reconfiguration port
    .vauxp12(xa_p[1]),
    .vauxn12(xa_n[1]),
    .vauxp4(xa_p[0]),
    .vauxn4(xa_n[0]),
    .busy_out(),                    // ADC Busy signal
    .channel_out(),                 // Channel Selection Outputs
    .do_out(ADC_data),              // Output data bus for dynamic reconfiguration port
    .drdy_out(ready),               // Data ready signal for the dynamic reconfiguration port
    .eoc_out(enable),               // End of Conversion Signal
    .vp_in(vp_in),                  // Dedicated Analog Input Pair
    .vn_in(vn_in)
);

//Read XADC analog input value and store into "Segment_Data" via shift register to create 32 bit seed value
always @(posedge CLK10Hz or negedge rstn) begin
if(!rstn)begin
    Segment_data<=0;
  end
  else begin
    Segment_data<=Segment_data<<16;
    Segment_data[15:0] <= ADC_data;
  end
end

reg [31:0] temp;
reg [31:0] temp2;
reg [31:0] rand;

//Pseudo-Random XOR shift register algorithm with seed value supplied by XADC, it's not perfect as seed keeps changing every number
always@(sysclk)begin
temp = Segment_data ^ Segment_data >>7;
temp2 = temp ^ temp << 9;
rand = temp2 ^ temp2 >> 13;
end

//setup for tilt detection and dice selection function
reg [15:0] out;
reg [3:0] ones_set;
reg [3:0] tens_set;
reg [3:0] huns_set;
reg [3:0] thou_set;
reg upright;
reg setmode;
reg [3:0] dselect;
reg [9:0] tiltlog;
reg [4:0] sumtilt;
reg [6:0] diceval;
assign dp = upright;
reg btnUr,btnDr;

//this function polls the tilt sensor every 1/10th of a second and adds it to a 10 bit shift register, it adds the total value of the register to the "sumtilt" variable, if a critical number of bits are 1 (meaning the dice unit has been upright and stable
//for enough time, sumtilt>7 aka, the dice was upright for 7/10th of the second) then it sets the upright register to 1, otherwise it sets it to 0.
always@(posedge CLK10Hz or negedge rstn)begin
  if(!rstn)begin
    sumtilt<=5'b00000;
```

```
    tiltlog<=9'b000000000;
    upright<=0;
  end
  else
  begin
    tiltlog <= tiltlog << 1;
    tiltlog[0]<=tilt;
sumtilt=tiltlog[9]+tiltlog[8]+tiltlog[7]+tiltlog[6]+tiltlog[5]+tiltlog[4]+tiltlog[3]+tiltlog[2]+tiltlog[1]+tiltlog[0];
  if (sumtilt>=7) begin
      upright<=1;
  end
  else begin
      upright<=0;
  end
  end
end

//This function detects when the set dice value buttons are pressed only when the dice is upright, and changes the dice's mode into dice set mode (setmode). pressing the buttons in this mode will allow the value of the dice (diceval) to be
//changed between 2,4,6,8,10,12,20,100 sided dice. The keep awake disable is also here, which requires both up and down buttons to be pressed at the same time.
always@(posedge CLK10Hz or negedge rstn)begin
  if(!rstn)begin
  thou_set<=0;
  huns_set<=0;
  tens_set<=0;
  ones_set<=0;
    btnUr<=0;
    btnDr<=0;
    dselect<=0;
    setmode<=0;
    diceval<=2;
    keepon<=1;
  end
    else begin

  if(btnU)
    btnUr<=1;
  else
    btnUr<=0;
  if(btnD)
    btnDr<=1;
  else
    btnDr<=0;

  if(upright)begin
      if(btnUr)begin
        if(btnDr)begin
            keepon<=0;
        end
        else begin
        setmode<=1;
        if(dselect==7)
          dselect<=0;
        else
          dselect<=dselect+1;
        end
        end
      else if(btnDr)begin
        setmode<=1;
        if(dselect==0)
          dselect<=7;
        else
          dselect<=dselect-1;
        end

        thou_set<=4'hd;
        case(dselect)
            4'd0:begin
            huns_set<=2;
            tens_set<=4'hf;
```

```
            ones_set<=4'hf;
            diceval<=2;
            end
            4'd1:begin
            huns_set<=4;
            tens_set<=4'hf;
            ones_set<=4'hf;
            diceval<=4;
            end
            4'd2:begin
            huns_set<=6;
            tens_set<=4'hf;
            ones_set<=4'hf;
            diceval<=6;
            end
            4'd3:begin
            huns_set<=8;
            tens_set<=4'hf;
            ones_set<=4'hf;
            diceval<=8;
            end
            4'd4:begin
            huns_set<=1;
            tens_set<=0;
            ones_set<=4'hf;
            diceval<=10;
            end
            4'd5:begin
            huns_set<=1;
            tens_set<=2;
            ones_set<=4'hf;
            diceval<=12;
            end
            4'd6:begin
            huns_set<=2;
            tens_set<=0;
            ones_set<=4'hf;
            diceval<=20;
            end
            4'd7:begin
            huns_set<=1;
            tens_set<=0;
            ones_set<=0;
            diceval<=100;
            end
         endcase

   end
    else
        setmode<=0;

end
end
// setup for the raw RNG value to be processed
reg [3:0] ones_rand;
reg [3:0] tens_rand;
reg [3:0] huns_rand;
reg [3:0] thou_rand;
reg [3:0] ones_randr;
reg [3:0] tens_randr;
reg [3:0] huns_randr;
reg [3:0] thou_randr;

//The raw RNG value is processed every 10th of a second, the value is modulo ed by the dice value to normalise the extended 32 bit range to the required dice range (eg, compressed from 0-4,294,967,295 to 1-6 for a 6 sided dice) the result is then converted to
//a BCD value for display, values are shifted by 10 to centralise the values displayed. The continious updating gives the "scambled" or "rolling" effect that scrolls through the values on the display.
always @(posedge CLK10Hz or negedge rstn) begin
if(!rstn)begin
    ones_rand=4'hd;
    tens_rand=4'hd;
    huns_rand=4'hd;
    thou_rand=4'hd;
```

```
  end
  else if(!upright) begin
    out=(rand%diceval)+1;   //32 bit to dice range conversion
    ones_randr=4'hf;        //ones segment disabled
    tens_randr = out % 10;  // Units digit displayed in tens segment
    out = out / 10;         // divisor fed to next modulo
    huns_randr = out % 10;  // Tens digit displayed in hundreds segment
    out = out / 10;         // divisor fed to next modulo
    thou_randr = out % 10;  // Hundreds digitdisplayed in thousands segment

    thou_rand=thou_randr;
    huns_rand=huns_randr;
    tens_rand=tens_randr;
    ones_rand=ones_randr;
    end
    else begin
    thou_rand=thou_rand; //if dice is upright, the values are held so they can be read
    huns_rand=huns_rand;
    tens_rand=tens_rand;
    ones_rand=ones_rand;

    end
  end

  //The keep awake function is needed as the battery charge/discharge module turns off after 30 seconds if current less than 50ma is drawn (the FPGA module current draw tends to dip below 50ma at times, thus it can sporadically shut off.
  //the button on the charge/discharge module is a on/off button, if the button is pressed in quick succession, it turns off, but if it's pressed after a long interval, it will refresh the 30 second shutdown timer.
  //the button is triggered by a falling edge pulled to ground.
  //this function prevents shutdown from happening by pulling the pin low every 10 seconds (5 sec clock) unless the keep awake is switched off by pressing both up and down buttons simultaneously
always @ (posedge CLK5s)begin
if(!rstn) begin
    onsig<=1;
    clk5<=0;
    end
else if(keepon) begin
    onsig<=~onsig;
    clk5<=~clk5;
    end
else begin
    onsig<=0;
    clk5<=0;
    end
end
assign onpin = onsig;

// setup to write values into segment display buffer
reg [3:0] ones_bcd;
reg [3:0] tens_bcd;
reg [3:0] huns_bcd;
reg [3:0] thou_bcd;
// selects values to be displayed from the dice selection mode or RNG mode based on the state it is in using the value (setmode)
always@(sysclk)begin
if(setmode)begin
thou_bcd<=thou_set;
huns_bcd<=huns_set;
tens_bcd<=tens_set;
ones_bcd<=ones_set;
end
else begin
if(thou_rand==0)
thou_bcd<=4'hf;
else
thou_bcd<=thou_rand;
if(huns_rand==0 && thou_rand==0)
huns_bcd<=4'hf;
else
huns_bcd<=huns_rand;
if(tens_rand==0 && huns_rand==00 && thou_rand ==0)
tens_bcd<=4'hf;
```

```
else
tens_bcd<=tens_rand;
if(ones_rand==0)
ones_bcd<=4'hf;
else
ones_bcd<=ones_rand;
end
end

//writes to registers to transfer to segment display function
assign  bcd_tim[15:12]  = thou_bcd;
assign  bcd_tim[11:8]   = huns_bcd;
assign  bcd_tim[7:4]    = tens_bcd;
assign  bcd_tim[3:0]    = ones_bcd;

//segment display function
Segment segment_u0(rstn,CLK500Hz,bcd_tim,{an[0],an[1],an[2],an[3]},seg[6:0]);

//uart setup
reg uart_ready;
wire uart_vaild;

//uart control function, this determines the rate at which messages are sent, not to be confused by bitrate.
//has some issues sending messages at a rate lower than 100Hz even though the actual data rate capable by this system is 100Hz
always@(posedge CLK1000Hz,negedge rstn)begin
    if (!rstn)begin
        uart_ready <= 1'b0;
    end
    else begin
uart_ready <= ~uart_ready;
    end
end

//outputs the value of the 100s and 10s segments (aka the tens and ones value of the diceroll) through UART for logging in the computer
//the signal is passed through the uart_rxd_out pin which is tied to the FT2232HQ USB-UART bridge to enable communication with the computer rather than needing to have a separate device to act as the middleman.
uart_tx uart_tx_u0(CLK1000Hz,rstn,uart_ready,uart_valid,uart_rxd_out,1'b0,{huns_rand,tens_rand});

endmodule
```